\documentclass[aps,twocolumn,showpacs]{revtex4}
\usepackage{graphics}
\usepackage{graphicx}

\renewcommand{\ss}{\scriptscriptstyle}
\input{tcilatex}

\begin{document}

\title{Transient quantum evolution of 2D electrons under photoexcitation of
a deep center}
\author{F.T. Vasko}
\email{ftvasko@yahoo.com}
\affiliation{Institute of Semiconductor Physics, NAS Ukraine, Pr. Nauki 41, Kiev, 03028,
Ukraine}
\author{A. Hernandez-Cabrera }
\email{ajhernan@ull.es}
\author{P. Aceituno}
\affiliation{Dpto. Fisica Basica, Universidad de La Laguna, La Laguna, 38206-Tenerife,
Spain}
\date{\today}

\begin{abstract}
We have considered the ballistic propagation of the 2D electron Wigner
distribution, which is excited by an ultrashort optical pulse from a
short-range impurity into the first quantized subband of a selectively-doped
heterostructure with high mobility. Transient ionization of a deep local
state into a continuum conduction $c$-band state is described. Since the
quantum nature of the photoexcitation, the Wigner distribution over 2D plane
appears to be an alternating-sign function. Due to a negative contribution
to the Wigner function, the mean values (concentration, energy, and flow)
demonstrate an oscillating transient evolution in contrast to the diffusive
classical regime of propagation.
\end{abstract}

\maketitle

\section{Introduction}

In recent decades, intensive efforts were paid in order to study the quantum
coherent properties of different physical systems \cite{1x}. During the
development of the ultrafast spectroscopy of bulk semiconductors and
heterostructures \cite{1}, both coherent oscillations between coupled states
and different relaxation processes have been investigated (see references in 
\cite{2} and \cite{3}). Some quantum peculiarities, e.g. in the transport of
mesoscopic devices \cite{3x} or in the dynamics of electron excited at
metallic surfaces \cite{3y}, were discussed recently but, to the best of our
knowledge, the coherent dynamics of a free quasiparticle, which propagates
over continuum states, was not measured directly in any solid state system.
The quantum response, such as the formation of quasiparticles in different
systems \cite{4,5,6,7}, has been observed for sub-picosecond stage of
evolution. Under theoretical consideration of such kind of measurements
(e.g., see \cite{8} and references therein), one can model the
photogeneration process using a simple initial condition describing the
creation of carriers during a femtosecond temporal interval. At the same
time, in the case of photoexcitation of carriers with low concentration and
with energy values below the optical phonon energy, the dynamical regime of
the response appears to be valid up to nanosecond time interval. It is
because both the fast relaxation, due to optical phonon emission, and the
carrier-carrier interaction are suppressed. Thus, a possibility is to study
the quantum nature of the ballistic transient evolution, caused by the
non-classical character of photoexcitation \cite{3, 9}.

Modern high-mobility heterostructures are characterized by a momentum
relaxation time correspondent to the subnanosecond scale at low temperatures 
\cite{10}. So that the mean free path appears to be macroscopic ($>$100 $\mu
m$ if electron energy is about few meV). The photoexcitation of a single
deep impurity under a laser pumping focused up to submicron scale \cite{11}
can be carried out in a non-doped heterostructure with a low surface
concentration of centers (deep centers in bulk GaAs are under consideration
since the starting of 70s \cite{12}). Below we consider the transient
photoexcitation of electrons from a deep impurity level and the quantum
ballistic evolution of the Wigner distribution in the 2D plane over
submillimeter distances during nanosecond time interval.

In contrast to transitions between local states, when the Rabi effect
(oscillations of population versus pumping intensity \cite{3, 9}) takes
place, the level population of a local state under ultrafast photoionization
decreases monotonically with the pumping intensity because the excited
electron appears to be delocalized over the conduction $c$-band. Another
peculiarity of the process under consideration is the quantum character of
the transient evolution. Due to this, the concentration distribution, which
decreases from the center, involves an oscillating contribution and regions
of \textit{a negative Wigner distribution} take place. Such a distribution
should be considered with the use of the quantum kinetic equation, written
in the Wigner representation, due to the following reasons: $(a)$ the energy
conservation law is not valid during the photoexcitation process and $(b)$
there is no momentum restrictions on the excited distribution due to the
short-range impurity state involved in the phototransition.

In this paper, we restrict ourself to the local time approximation which
corresponds to the photoionization above the $c$-band edge ($\Delta \omega
\tau _{ex}>1$, where $\Delta \omega $ is the detuning frequency and $\tau
_{ex}$ is the duration of photoexcitation), when only the point $(b)$ is
essential. Due to this reason, the mean values (concentration, energy, and
flow) show an oscillating behavior in contrast to the diffusive classical
regime. Moreover, although the concentration and energy distributions are
positive-definite functions, the flow distribution appears to be an
alternating-sign one, i.e. a flow may be directed opposite to a
concentration gradient. The peculiarities discussed can be verified by the
use of optical methods or scanning tunneling microscopy, if the measurements
can be performed with submicron and subnanosecond resolutions.

The present work is organized as follows. The photoexcitation process,
including the evolution of the deep center population and the transient
Wigner distribution over $c$-band, is described in Sec. II. Section III
presents temporal dependencies of the above-introduced functions. The
transient dynamics of the mean values is described in Sec. IV. A list of the
assumptions used and the discussion of the methods for experimental
verification of the peculiarities discussed are given in the concluding
section. Appendix contains the description of the classical regime of
transient evolution.

\section{Ultrafast Photoexcitation}

Under photoexcitation of electrons, transitions from a deep local level into
the first subband of $c$-band is described by the density matrix of the $j$%
-state, $\widehat{\rho }_{jt}$. Performing the averaging over the period of
the radiation $\mathbf{E}_{t}\exp (-i\omega t)+c.c.$ one obtains the quantum
kinetic equation \cite{9}: %1
\begin{equation}
\frac{\partial \widehat{\rho }_{jt}}{\partial t}+\frac{i}{\hbar }\left[ 
\widehat{h}_{j},\widehat{\rho }_{jt}\right] =\widehat{G}_{jt}
\end{equation}%
with the generation rate ($j\neq j^{\prime }$) %2
\begin{eqnarray}
\widehat{G}_{jt} = \left( \frac{e}{\hbar \omega }\right) ^{2}\int_{-\infty
}^{t}dt^{\prime }e^{-i\omega (t^{\prime }-t)}~~~~  \nonumber \\
\times \left\{ \hat{S}_{jt^{\prime }-t}^{+}\left( \mathbf{E}_{t^{\prime
}}\cdot \widehat{\mathbf{v}}\right) _{jj^{\prime }}\widehat{\rho }%
_{j^{\prime }t^{\prime }}\hat{S}_{j^{\prime }t^{\prime }-t}\left( \mathbf{E}%
_{t}\cdot \widehat{\mathbf{v}}^{+}\right) _{j^{\prime }j}\right. ~~~ 
\nonumber \\
\left. +\left( \mathbf{E}_{t}\cdot \widehat{\mathbf{v}}^{+}\right)
_{jj^{\prime }}\hat{S}_{j^{\prime }t^{\prime }-t}^{+}\widehat{\rho }%
_{j^{\prime }t^{\prime }}\left( \mathbf{E}_{t^{\prime }}\cdot \widehat{%
\mathbf{v}}\right) _{j^{\prime }j}\hat{S}_{jt^{\prime }-t}\right\} _{j\neq
j^{\prime }}+H.c..
\end{eqnarray}%
Here $\widehat{h}_{j}$ is the Hamiltonian of the $2D$ state in the $c$-band (%
$j=c$) or of the state at short-range centre ($j=h$), $\hat{S}_{jt^{\prime
}-t}=\exp [-i\hat{h}_{j}(t^{\prime }-t)/\hbar ]$ is the evolution operator
of the $j$-th state, and $(\widehat{\mathbf{v}})_{jj^{\prime }}$ is the
velocity matrix element for $j\leftrightarrow j^{\prime }$ transitions. For
the case of a deep center connected to the valence $v$-band \cite{13}, we
use in (2) the interband matrix element of velocity, $\mathrm{v}_{cv}$,
multiplied by the overlap integral between the plane wave of momentum $%
\mathbf{p}$ and the local state, $I_{\mathbf{p}}=\langle \mathbf{p}|h\rangle 
$. The evolution of the distribution function over $c$-band, $f_{\mathbf{p}%
_{1}\mathbf{p}_{2}t}=\langle \mathbf{p}_{1}|\widehat{\rho }_{ct}|\mathbf{p}%
_{2}\rangle $, is governed by the equation: %3
\begin{eqnarray}
\frac{\partial f_{\mathbf{p}_{1}\mathbf{p}_{2}t}}{\partial t}+\frac{i}{\hbar 
}(\varepsilon _{p_{1}}-\varepsilon _{p_{2}})f_{\mathbf{p}_{1}\mathbf{p}%
_{2}t} = G_{\mathbf{p}_{1}\mathbf{p}_{2}t},  \nonumber \\
G_{\mathbf{p}_{1}\mathbf{p}_{2}t} \simeq \left( \frac{eE\mathrm{v}_{cv}}{%
\hbar \omega }\right) ^{2}I_{\mathbf{p}_{1}}I_{\mathbf{p}_{2}}^{\ast
}w_{t}\int_{-\infty }^{t}dt^{\prime }w_{t^{\prime }}  \nonumber \\
\times e^{i(\varepsilon _{p}/\hbar -\Delta \omega )(t^{\prime
}-t)}n_{t^{\prime }}+(c.c.,\mathbf{p}_{1} \longleftrightarrow \mathbf{p}_{2})
\end{eqnarray}%
with the right-hand side dependent on the population of the local state, $%
n_{t}\equiv \langle h|\widehat{\rho }_{ht}|h\rangle $. The generation rate, $%
G_{\mathbf{p}_{1}\mathbf{p}_{2}t}$, is determined through the $2D$ kinetic
energy $\varepsilon _{p}=p^{2}/m$ with the effective mass of $c$-band, $m$,
the form-factor $w_{t}$ introduced by the relation $\mathbf{E}_{t}=\mathbf{E}%
w_{t}$, and the detuning energy $\hbar \Delta \omega ,$ which takes into
account the subband quantization.

We are using the initial conditions $f_{\mathbf{p}_{1}\mathbf{p}%
_{2}t\rightarrow -\infty }=0$ and $n_{t\rightarrow -\infty }=1$, which
correspond to the single-electron population of the spin-degenerated level,
so that the normalization condition takes the form: $n_{t}+2\sum_{\mathbf{p}%
}f_{\mathbf{p,p}t}=1$. Evolution of the local state population is governed
by the integro-differential equation %4
\begin{eqnarray}
\frac{dn_{t}}{dt}+2\left( \frac{eE\mathrm{v}_{cv}}{\hbar \omega }\right)
^{2}w_{t}\int_{-\infty }^{t}dt^{\prime }w_{t^{\prime }}n_{t^{\prime }} 
\nonumber \\
\times \sum_{\mathbf{p}}|I_{\mathbf{p}}|^{2}\cos \left( \frac{\varepsilon
_{p}}{\hbar }-\Delta \omega \right) (t-t^{\prime }) = 0,
\end{eqnarray}%
which is obtained from Eqs. (1, 2). Instead of Eq. (3), one can describe the
transient evolution of the $c$-band distribution through the Wigner
function, $f_{\mathbf{p,q}t}\equiv f_{\mathbf{p}-\hbar \mathbf{q}/2,\mathbf{p%
}+\hbar \mathbf{q}/2t}$ which is governed by the equation %5
\begin{equation}
\left( \frac{\partial }{\partial t}+i\mathbf{q}\cdot \mathbf{v}\right) f_{%
\mathbf{p,q}t}=G_{\mathbf{p,q}t},
\end{equation}%
with the velocity $\mathbf{v}=\mathbf{p}/m$. Similarly, the generation rate, 
$G_{\mathbf{p,q}t}\equiv G_{\mathbf{p}-\hbar \mathbf{q}/2,\mathbf{p}+\hbar 
\mathbf{q}/2t}$, is transformed into %6
\begin{eqnarray}
G_{\mathbf{p,q}t} \simeq \left( \frac{eE\mathrm{v}_{cv}}{\hbar \omega }%
\right) ^{2}I_{\mathbf{p}+\frac{\hbar \mathbf{q}}{2}}I_{\mathbf{p}-\frac{%
\hbar \mathbf{q}}{2}}w_{t}\int_{-\infty }^{t}dt^{\prime }w_{t^{\prime
}}n_{t^{\prime }}~~~  \nonumber \\
\times \left\{ \exp \left[ i\left( \frac{\varepsilon _{\mathbf{p}+\frac{%
\hbar \mathbf{q}}{2}}}{\hbar }-\Delta \omega \right) (t^{\prime }-t)\right]
+(c.c.,\mathbf{q}\rightarrow -\mathbf{q})\right\}
\end{eqnarray}%
and the right-hand side of Eq. (5) is determined through the evolution of $%
n_{t}$. The solution of (5) takes the form $f_{\mathbf{p,q}t}=\int_{-\infty
}^{t}dt^{\prime }\exp [-i\mathbf{q\cdot v}(t-t^{\prime })]G_{\mathbf{p,q}%
t^{\prime }}$, so the description of the transient evolution is reduced to
the calculation of a multiple integral and to the solution of Eq. (4).

A simplified consideration of the problem is possible under the condition $%
\Delta \omega \tau _{ex}>1$ (photoionization into a high-energy state of $c$%
-band) when $w_{t^{\prime }}n_{t^{\prime }}$ in Eqs. (4) and (6) can be
replaced by $w_{t}n_{t}$ due to the fast oscillating factors (the local time
approximation). The integration over $dt^{\prime }$ in Eq. (6), which is
performed with an infinitesimal damping factor in the exponent, $\delta
\rightarrow +0$, gives %7
\begin{eqnarray}
G_{\mathbf{p,q}t} \approx \left( \frac{eE\mathrm{v}_{cv}}{\hbar \omega }%
\right) ^{2}I_{\mathbf{p}+\hbar \mathbf{q}/2}I_{\mathbf{p}-\hbar \mathbf{q}%
/2}w_{t}^{2}n_{t}~~~  \nonumber \\
\times \left[ \frac{\hbar /i}{\varepsilon _{\mathbf{p}+\hbar \mathbf{q}%
/2}-\hbar \Delta \omega -i\delta }+(c.c.,\mathbf{q}\rightarrow -\mathbf{q})%
\right] .
\end{eqnarray}%
Here the kinetic energy of the inhomogeneous system ($\mathbf{q}\neq 0$) is
not conserved during the photogeneration process, even for the long $\tau
_{ex}$ case, due to the violation of the momentum conservation law. Using
Eq. (7) and performing the Fourier transformation of the distribution
function $f_{\mathbf{p,q}t}$ one obtains: %8
\begin{eqnarray}
f_{\mathbf{p,x}t} = \sum_{\mathbf{q}}\int_{-\infty }^{t}dt^{\prime }e^{i%
\mathbf{q\cdot x}_{t-t^{\prime }}}G_{\mathbf{p,q}t^{\prime }}~~~~~~~ 
\nonumber \\
\approx 2\hbar \left( \frac{eE\mathrm{v}_{cv}}{\hbar \omega }\right)
^{2}\int_{-\infty }^{t}dt^{\prime }w_{t^{\prime }}^{2}n_{t^{\prime }}\sum_{%
\mathbf{q}}I_{\mathbf{p}+\frac{\hbar \mathbf{q}}{2}}I_{\mathbf{p}-\frac{%
\hbar \mathbf{q}}{2}}  \nonumber \\
\times \left[ \pi \delta (\varepsilon _{\mathbf{p}+\frac{\hbar \mathbf{q}}{2}%
}-\hbar \Delta \omega )\cos (\mathbf{q\cdot x}_{t-t^{\prime }})+\frac{\sin (%
\mathbf{q\cdot x}_{t-t^{\prime }})}{\varepsilon _{\mathbf{p}+\frac{\hbar 
\mathbf{q}}{2}}-\hbar \Delta \omega }\right] ,
\end{eqnarray}%
where we have introduced the time-dependent coordinate, $\mathbf{x}%
_{t-t^{\prime }}=\mathbf{x}-\mathbf{v}(t-t^{\prime })$. The function $f_{%
\mathbf{p,x}t}$ satisfies the conditions $f_{\mathbf{-p,-x}t}=f_{\mathbf{p,x}%
t}$ and $f_{\mathbf{-p,x}t}=f_{\mathbf{p,-x}t}$, which are verified by the
exchange $\mathbf{q}\rightarrow -\mathbf{q}$ in Eq. (8).

Within the local time approximation, Eq. (4) takes the form: %9
\begin{eqnarray}
\left( \frac{d}{dt}+\gamma w_{t}^{2}\right) n_{t} =0,~~~~~~  \nonumber \\
\gamma =4\pi \left( \frac{eE\mathrm{v}_{cv}}{\hbar \omega }\right) ^{2}\sum_{%
\mathbf{p}}|I_{\mathbf{p}}|^{2}\delta \left( \varepsilon _{p}/\hbar -\Delta
\omega \right) ,
\end{eqnarray}%
where $\gamma $ stands for the photoionization decrement. The analytical
solution of Eq. (9), %10
\begin{equation}
n_{t}=1-\gamma \int_{-\infty }^{t}dt^{\prime }w_{t^{\prime }}^{2}\exp \left(
-\gamma \int_{t^{\prime }}^{t}dt^{\prime \prime }w_{t^{\prime \prime
}}^{2}\right) ,
\end{equation}%
describes the transient population of the level \cite{13x}. Thus, we have
obtained the Wigner distribution (8), and the population (10) written in the
integral forms which contain the overlap integral $I_{\mathbf{p}}$ and the
form-factor $w_{t}$.

\begin{figure}[tbp]
\begin{center}
\includegraphics{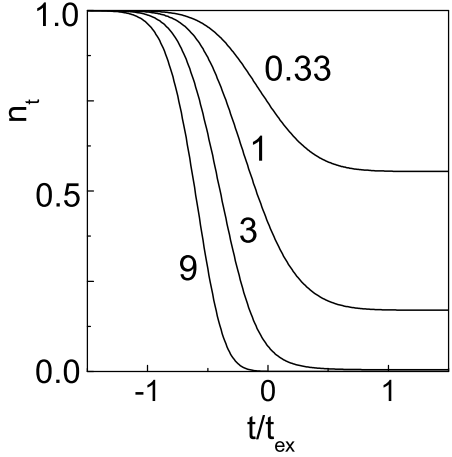}
\end{center}
\par
\addvspace{-0.7 cm}
\caption{Transient evolution of the population of short-range level under
the dimensionless pumpings $\protect\gamma \protect\tau _{ex}/2=$0.33, 1, 3
and 9.}
\end{figure}

\section{Short-range case}

To calculate the distribution (8) and the population (10) we use below the
overlap integral for the short-range local state with the characteristic
size $l_{o}$, so that $I_{\mathbf{p}}\simeq l_{o}/L$ if $p<\hbar /l_{o}$ and 
$I_{\mathbf{p}}\simeq 0$ if $p>\hbar /l_{o}$; here $L$ is the normalization
length. Within the above assumption, the decrement of photoionization in Eq.
(9) takes the form: %11
\begin{equation}
\gamma \simeq \left( \pi \frac{eE\mathrm{v}_{cv}}{\hbar \omega }l_{o}\right)
^{2}\hbar \rho _{\ss 2D}=\frac{\pi ^{3}m}{2m_{h}}\left( \frac{eE\mathrm{v}%
_{cv}}{\hbar \omega }\right) ^{2}\frac{\hbar }{\Delta E},
\end{equation}%
where $\rho _{\ss 2D}$ is the density of states, $m_{h}$ is the heavy hole
mass, and the level coupling energy, $\Delta E$, is expressed through $l_{o}$
according to $\Delta E\simeq (\hbar /l_{o})^{2}/2m_{h}$. The temporal
dependencies of $n_{t}$ under different pumping level, which is determined
by the dimensionless parameter $\gamma \tau _{ex}/2$, are shown in Fig. 1
for the Gaussian form-factor $w_{t}=\exp [-2(t/\tau _{ex})^{2}]$. The
complete ionization of the center appears under the condition $\gamma \tau
_{ex}/2\sim $2 and, when $\gamma $ increases, the photoionization takes
place during the front of pulse. The full ionization regime takes place
under a pulse energy $\sim 0.2\mu J$ focused on an area $\sim 100\mu m$;
this estimate is performed for the GaAs parameters and does not depend on
the pulse duration.

Next, we turn to the description of the photoexcited electron distribution
given by Eq. (8) and dependent on time, $|\mathbf{p}|$, $|\mathbf{x}|$, and
the angle $\widehat{\mathbf{p,x}}$. We consider the long-duration excitation
case (the dynamic regime of response takes place for the nanosecond time
scale) and demonstrate that the Wigner distribution $f_{\mathbf{p,x}t}$ is
not a positive-definite function. We calculate below the distribution at the
maximal pumping, $t=0$, for the cases $\mathbf{p}\Vert \mathbf{x}$ and $%
\mathbf{p}\bot \mathbf{x}$ with the use of the notations $f_{p,x}^{\ss \Vert
}$ and $f_{p,x}^{\ss \perp }$, respectively. Performing in Eq. (8) the
integration over $\mathbf{q}\bot \mathbf{p}$ one obtains the distributions $%
f_{p,x}^{\ss \Vert ,\perp }$ as follows: %12
\begin{eqnarray}
\left\vert 
\begin{array}{c}
f_{p,x}^{\ss \Vert } \\ 
f_{p,x}^{\ss \perp }%
\end{array}%
\right\vert = \frac{2\gamma }{\pi }\int_{-\infty
}^{0}dtw_{t}^{2}n_{t}\int_{-\infty }^{\infty }dp_{1}\left\{ \frac{\theta
\lbrack p_{\Delta \omega }^{2}-p_{1}^{2}]}{\sqrt{p_{\Delta \omega
}^{2}-p_{1}^{2}}}\right.  \nonumber \\
\times \left\vert 
\begin{array}{c}
\cos \left[ \frac{2(p_{1}-p)}{\hbar }(x+vt)\right] \\ 
\cos \left[ \frac{2x}{\hbar }\sqrt{p_{\Delta \omega }^{2}-p_{1}^{2}}+\frac{%
2(p_{1}-p)}{\hbar }vt\right]%
\end{array}%
\right\vert +\frac{\theta \lbrack p_{1}^{2}-p_{\Delta \omega }^{2}]}{\sqrt{%
p_{1}^{2}-p_{\Delta \omega }^{2}}}~~~~~  \nonumber \\
\left. \times \left\vert 
\begin{array}{c}
\sin \left[ \frac{2(p_{1}-p)}{\hbar }(x+vt)\right] \\ 
\exp \left[ -\frac{2x}{\hbar }\sqrt{p_{1}^{2}-p_{\Delta \omega }^{2}}\right]
\sin \left[ \frac{2(p_{1}-p)}{\hbar }vt\right]%
\end{array}%
\right\vert \right\} ,
\end{eqnarray}%
where $\theta \lbrack z]$ is the Heaviside step function, $v=|\mathbf{v}|$,
and $p_{\Delta \omega }=\sqrt{2m\hbar \Delta \omega }$ is the characteristic
momentum. The integrals over $t$ and $p_{1}$ can be factorized for the slow
electron case, $p\simeq 0$; moreover, a non-zero contribution appears from
the first addendum only. The distributions for the $\Vert $ and $\perp $
orientations are coincident, %13
\begin{equation}
f_{p=0,x}^{\ss \Vert ,\perp }\simeq 2\gamma \int_{-\infty
}^{0}dtw_{t}^{2}n_{t}J_{0}\left( \frac{2p_{\Delta \omega }x}{\hbar }\right) ,
\end{equation}%
and the coordinate dependence is given by the zero-order Bessel function, $%
J_{0}(z)$, which has an alternating-sign value and decreases as a square
root.

The distribution functions (12) depend on the dimensionless momentum and
coordinate, $p/p_{\Delta \omega }$ and $xp_{\Delta \omega }/\hbar $, the
detuning parameter, $\Delta \omega \tau _{ex}$, and the pumping intensity, $%
\gamma $. Using (10) with the dimensionless pumping $\gamma \tau _{ex}/2=3$
and performing the numerical integration in Eq. (12) we plot the Wigner
distribution for $\Delta \omega \tau _{ex}=10$ as it is shown in Fig. 2. One
can see a non-monotonically dependency on $p/p_{\Delta \omega }$ and $%
xp_{\Delta \omega }/\hbar $ with pronounced negative contributions. Both
longitudinal and transverse cases show a fast decrease with dimensionless
momentum, whereas oscillations slowly decrease with dimensionless coordinate
due to the spread of the distribution under propagation.

\begin{figure}[tbp]
\begin{center}
\includegraphics{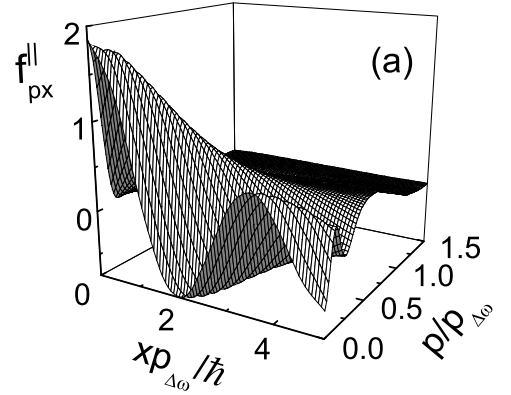} \includegraphics{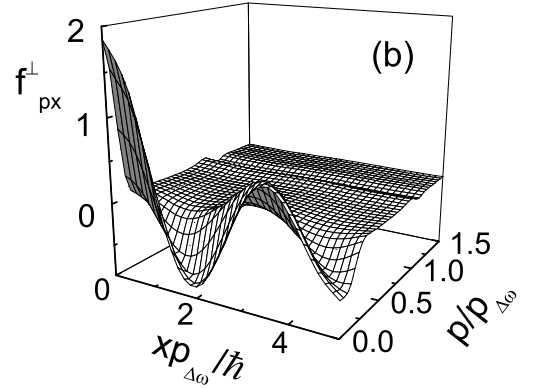}
\end{center}
\caption{Longitudinal $(a)$ and transverse $(b)$ Wigner distributions at
maximal pumping ($t=0$) versus dimensionless momentum and coordinate, $%
p/p_{\Delta \protect\omega }$ and $xp_{\Delta \protect\omega }/\hbar $.}
\end{figure}

\section{Mean values}

The transient dynamics of the Wigner distribution under consideration can be
verified by the treatment of spatio-temporal dependencies of the mean values
(concentration, energy, and flow, $n_{\mathbf{x}t}$, $\mathcal{E}_{\mathbf{x}%
t}$, and $\mathbf{i}_{\mathbf{x}t}$) given by the standard formulas: %14
\begin{equation}
\left\vert 
\begin{array}{c}
n_{\mathbf{x}t} \\ 
\mathcal{E}_{\mathbf{x}t} \\ 
\mathbf{i}_{\mathbf{x}t}%
\end{array}%
\right\vert =2\int \frac{d\mathbf{p}}{(2\pi \hbar )^{2}}\left\vert 
\begin{array}{c}
1 \\ 
\varepsilon _{p} \\ 
\mathbf{v}%
\end{array}%
\right\vert f_{\mathbf{px}t}.
\end{equation}%
Below we analyze the spatio-temporal evolution of (14) using the
distribution (8). Performing the integration over the variable $\mathbf{p}%
\pm \hbar \mathbf{q}/2$ one obtains the concentration and energy
distributions, which are isotropic over the $\mathbf{x}$-plane: %15
\begin{equation}
\left\vert 
\begin{array}{c}
n_{xt} \\ 
\mathcal{E}_{xt}%
\end{array}%
\right\vert =\frac{\gamma }{2\pi }\int_{-\infty }^{t}dt^{\prime
}w_{t^{\prime }}^{2}n_{t^{\prime }}\int_{0}^{\infty }dqqJ_{0}(qx)\left\vert 
\begin{array}{c}
N_{q,t-t^{\prime }} \\ 
E_{q,t-t^{\prime }}%
\end{array}%
\right\vert .
\end{equation}%
Here the kernels $N_{q,\tau }$ and $E_{q,\tau }$ are given by %16
\begin{eqnarray}
N_{q,\tau } = \cos \left( \frac{\varepsilon _{\hbar q}\tau }{\hbar }\right)
J_{0}(qv_{\Delta \omega }\tau )~~~  \nonumber \\
+\sin \left( \frac{\varepsilon _{\hbar q}\tau }{\hbar }\right) \mathcal{P}%
\int_{0}^{\infty }dy\frac{J_{0}(qv_{\Delta \omega }\tau \sqrt{y})}{\pi (y-1)}
\end{eqnarray}%
and %17
\begin{eqnarray}
E_{q,\tau } = \cos \left( \frac{\varepsilon _{\hbar q}\tau }{\hbar }\right)
\left( \hbar \Delta \omega +\frac{\varepsilon _{\hbar q}}{4}\right)
J_{0}(qv_{\Delta \omega }\tau )~~~  \nonumber \\
+\sin \left( \frac{\varepsilon _{\hbar q}\tau }{\hbar }\right) \mathcal{P}%
\int_{0}^{\infty }dy\left( \hbar \Delta \omega y+\frac{\varepsilon _{\hbar q}%
}{4}\right) \frac{J_{0}(qv_{\Delta \omega }\tau \sqrt{y})}{\pi (y-1)} 
\nonumber \\
+\hbar qv_{\ss \Delta \omega }\left[ \cos \left( \frac{\varepsilon _{\hbar
q}\tau }{\hbar }\right) \mathcal{P}\int_{0}^{\infty }dy\sqrt{y}\frac{%
J_{1}(qv_{\Delta \omega }\tau \sqrt{y})}{\pi (y-1)}\right.  \nonumber \\
\left. -\sin \left( \frac{\varepsilon _{\hbar q}\tau }{\hbar }\right)
J_{1}(qv_{\Delta \omega }\tau )\right] ~~~~~
\end{eqnarray}%
where $\mathcal{P}$ means the principal value of the integral, $J_{1}(z)$ is
the first-order Bessel function, $v_{\Delta \omega }\equiv p_{\Delta \omega
}/m$, and $y=\varepsilon /\hbar \Delta \omega $ is the dimensionless energy.
The distributions (15) depend on the pumping intensity through $n_{t},$
given by Eq. (10), and on the dimensionless coordinate and time, $xp_{\Delta
\omega }/\hbar $ and $t/\tau _{ex}$.

Before numerical calculations, we consider the asymptotes of $n_{xt}$
(similar formulas can be written for $\mathcal{E}_{xt}$ and the flow
distribution) for the case $t\gg \tau _{ex}$ and $x\gg \hbar /p_{\Delta
\omega }$. Using the asymptotic expansion of the Bessel function for large
arguments and performing the integrations over $t^{\prime }$ and $q$, one
obtains the explicit expression %18
\begin{eqnarray}
n_{xt} = \frac{\gamma \tau _{ex}\mathcal{N}}{(2\pi )^{2}l_{\Delta \omega }%
\sqrt{xv_{\Delta \omega }t}}\sqrt{\frac{\pi }{\Delta \omega t}}\left\{ \sin
\left( z_{-}^{2}+\frac{\pi }{4}\right) \right.  \nonumber \\
-\sin \left( z_{+}^{2}+\frac{\pi }{4}\right) +\mathcal{P}\int_{0}^{\infty }%
\frac{dy}{\pi \sqrt{y}(y-1)}  \nonumber \\
\left. \times \left[ \cos \left( z_{y-}^{2}+\frac{\pi }{4}\right) -\sin
\left( z_{y+}^{2}+\frac{\pi }{4}\right) \right] \right\} ,
\end{eqnarray}%
where $\mathcal{N}=\int_{-\infty }^{\infty }dtw_{t}^{2}n_{t}/\tau _{ex}$, $%
l_{\Delta \omega }=\hbar /p_{\Delta \omega }$, and we have introduced the
dimensionless forms $z_{\pm }=(x\pm v_{\Delta \omega }t)p_{\Delta \omega
}/2\hbar \sqrt{\Delta \omega t}$ and $z_{y\pm }=z_{\pm }\pm (\sqrt{y}-1)%
\sqrt{\Delta \omega t}$. Thus, the period of oscillations of the
concentration distribution, $(\hbar /2p_{\Delta \omega })\sqrt{\Delta \omega
t}$, does not depend on $\Delta \omega $ and increases as $\sqrt{t}$. 
\begin{figure}[tbp]
\begin{center}
\includegraphics{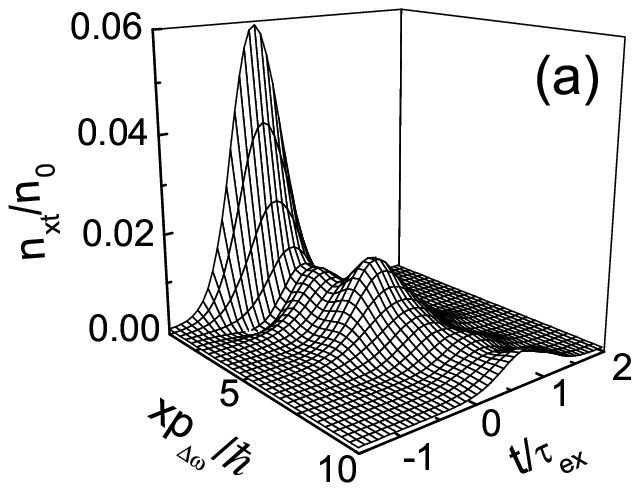} \includegraphics{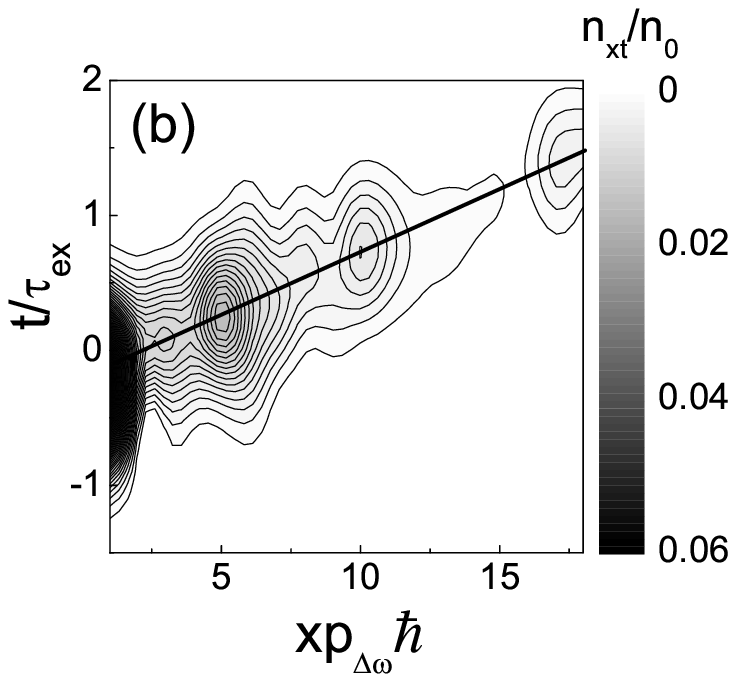}
\end{center}
\caption{Spatio-temporal evolution of concentration, $n_{\mathbf{x}t}$, for $%
\protect\gamma \protect\tau _{ex}/2=$ 1. $(a)$ 3D graph near the peak, and $%
(b)$ contour plot showing the line corresponding to the classical velocity.}
\end{figure}

The spatio-temporal dependency of the concentration is shown in Figs. 3.
Fig. 3$\left( a\right) $ shows the 3D graph near the peak at $xp_{\Delta
\omega }/\hbar =0$ for $\Delta \omega \tau _{ex}=5$ and $\gamma \tau
_{ex}/2=1$. As can be seen, concentration falls quickly for small
dimensionless coordinate and oscillates for bigger $xp_{\Delta \omega
}/\hbar $ values. At the same time\ maximum position in dimensionless time
is shifted following the classical velocity as shown in Fig. 3$(b)$ where
the contour plot together with the line corresponding to the classical
velocity is presented. Concentration is normalized by $n_{0}=\gamma \tau
_{ex}m\Delta \omega /\pi \hbar $, which is equal to $2.82\times 10^{11}$ cm$%
^{-2}$ for $\hbar \Delta \omega =5$ meV, and $\gamma \tau _{ex}/2=1$. This
value corresponds to an excited electron localized over an area of the order
of $(\hbar /p_{\Delta \omega })^{2}$. 
\begin{figure}[tbp]
\begin{center}
\includegraphics{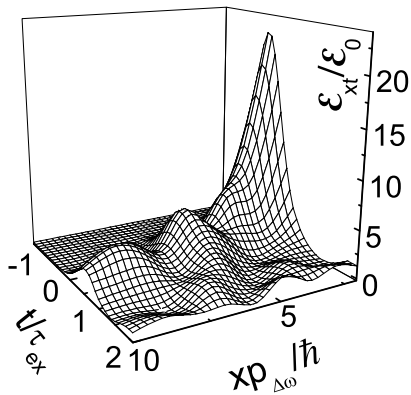}
\end{center}
\caption{Evolution of energy, $\mathcal{E}_{xt}$, for the same parameters of
Fig. 3.}
\end{figure}

Fig. 4 shows the behavior of the energy distribution vs. dimensionless
position and time. Energy distribution has been normalized by $%
E_{0}=n_{0}\hbar \Delta \omega $, which corresponds to $2.25\times 10^{-3}$
erg/cm$^{2}$ for the same values of $\hbar \Delta \omega $ and $\gamma \tau
_{ex}/2$ used for the concentration. Energy behavior vs. \ $xp_{\Delta
\omega }/\hbar $ is similar to the concentration one. To say, a fast
decrease followed by oscillations. 
\begin{figure}[tbp]
\begin{center}
\includegraphics{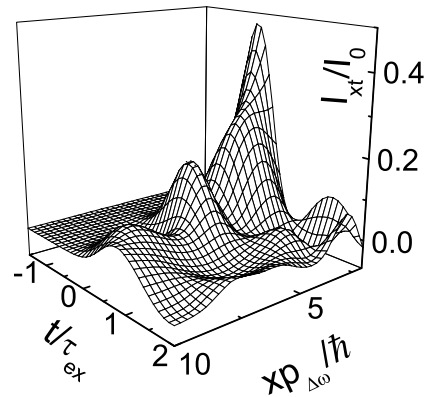}
\end{center}
\caption{Evolution of the flow, $I_{xt}$, for the same parameters of Fig. 3.}
\end{figure}

Since the in-plane isotropy of the problem, one obtains the flow density: $%
\mathbf{i}_{\mathbf{x}t}=(\mathbf{x}/|\mathbf{x}|)I_{xt}$, where the scalar
function $I_{xt}$ takes a similar form to (15): %19
\begin{equation}
I_{xt}=\frac{\gamma }{2\pi }\int_{-\infty }^{t}dt^{\prime }w_{t^{\prime
}}^{2}n_{t^{\prime }}\int_{0}^{\infty }dqqJ_{1}(qx)F_{q,t-t^{\prime }}
\end{equation}%
with the kernel %20
\begin{eqnarray}
F_{q,\tau } = \frac{\hbar q}{2m}\left[ \cos \left( \frac{\varepsilon _{\hbar
q}\tau }{\hbar }\right) \mathcal{P}\int_{0}^{\infty }dy\frac{%
J_{0}(qv_{\Delta \omega }\tau \sqrt{y})}{\pi (y-1)}\right. ~~~  \nonumber \\
\left. -\sin \left( \frac{\varepsilon _{\hbar q}\tau }{\hbar }\right)
J_{0}(qv_{\Delta \omega }\tau )\right] -v_{\ss \Delta \omega }\left[ \cos
\left( \frac{\varepsilon _{\hbar q}\tau }{\hbar }\right) J_{1}(qv_{\Delta
\omega }\tau )\right.  \nonumber \\
-\left. \sin \left( \frac{\varepsilon _{\hbar q}\tau }{\hbar }\right) 
\mathcal{P}\int_{0}^{\infty }\frac{dy\sqrt{y}}{\pi (y-1)}J_{1}(qv_{\Delta
\omega }\tau \sqrt{y})\right] .
\end{eqnarray}%
Performing the numerical integrations given by Eqs. (19, 20) we plot the
flow distribution for the above parameters, as it is shown in Fig. 5. Flow
distribution has also been normalized by $I_{0}=n_{0}p_{\Delta \omega }/m$,
being $I_{0}=4.57\times 10^{18}$ (cm/s)/cm$^{2}$ for the above values of $%
\hbar \Delta \omega $ and $\gamma \tau _{ex}/2$. Once again it appears an
initial fast decrease, followed by oscillations in dimensionless coordinate.
Moreover, as can be seen, there are oscillations also in time and negative
values of the flow distribution arise.

\section{Conclusions}

In summary, we have suggested a new scheme to investigate the quantum
peculiarities of the single-particle dynamics under ultrafast
photoionization of a single deep impurity. Due to the negative contributions
to transient Wigner distribution of $c$-band electron, the mean
concentration, energy, and flow demonstrate an oscillatory behavior in
contrast to the classical results (see Appendix). We have analyzed the
conditions for visible quantum oscillations, when a direct experimental
mapping of the quantum distribution should be possible.

Now we turn to the discussion of possibilities for experimental verification
of the peculiarities obtained. The stage of selective single-electron
photoexcitation is based on the assumption of a low concentration of deep
impurities: if a bulk concentration less than 10$^{12}$cm$^{-3}$ remains in
the near-surface region, one obtains an inter-center distance about 1$\mu $%
m. The regime of a single-center excitation can be easily realized with an
ultrafast pump focused over a submicron scale. Recently, similar
measurements were performed with a single quantum dot \cite{14} but the
photoexcitation into continuum and further evolution of distribution was not
examined. Perhaps, it is due to the complicate problem of the registration
of the oscillating Wigner distribution. In spite of the sensitive optical
methods developed recently for optical control of a single quantum dot (see 
\cite{15, 16} and Refs. therein) the spatial resolution remains a complicate
task (we use above $\hbar /p_{\Delta \omega }\sim $ $10$ nm). Note that the
period of the oscillations increases with time as $\sqrt{t}$ [see Eq. (18)
and Fig. 3$b$] but the distribution value (and the response) decreases due
to spatial spread. Another possibility is to use the scanning tunneling
microscopy \cite{17} which has nanometer resolution but has to be adapted to
time-resolved measurements with subnanosecond resolution. Note, that we do
not calculate any concrete optical or tunneling response supposing that the
observed peculiarities will be of the same order as the mean values
considered in Sec. IV.

Next, we discuss the assumptions used in our calculations. The main
approximation is the local time approach, so that the edge photoexcitation
is beyond of our consideration. A more complicate numerical simulation is
required for this case as well as to take into account the Coulomb
correlations (excitonic effect). The interaction of the electron with the
localized hole is essential for a near-center region but it should decrease
with $x$. Thus, the short-range model used in our calculations of Eqs. (9,
11) is enough in order to estimate the photoionization decrement. Finally,
only the averaged Wigner distribution has been considered and a full
counting statistics of photoionization and subsequent transient propagation
of electrons \cite{17} requires a special investigation.

In closing, a similar theoretical analysis may be developed for other cases
like photoexcitation of a single quantum dot or near-field photoexcitation 
\cite{11, 14}, where a similar quantum behavior should take place. We hope
that these results will stimulate experimental efforts towards a mapping of
quantum peculiarities in the transient Wigner distribution.

\begin{figure}[tbp]
\begin{center}
\includegraphics{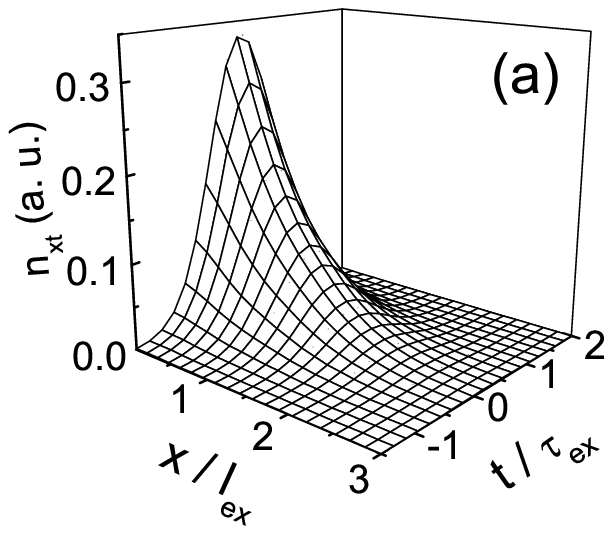} \includegraphics{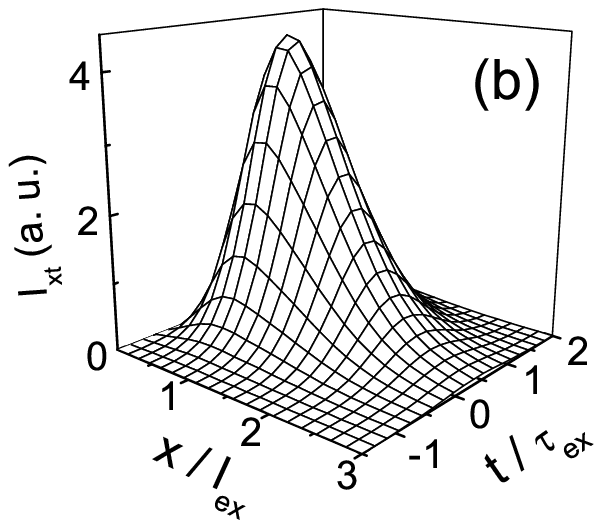}
\end{center}
\caption{Evolution of the classical distributions $n_{xt}$ $(a)$ and $I_{xt}$
given by Eqs. (A3) and (A4) for $\protect\alpha =$6.}
\end{figure}

\appendix

\section{Classical evolution}

This appendix contains the description of the classical regime of transient
evolution when the generation rate in Eq. (8) is approximated by the
factorized expression: %A1
\begin{equation}
G_{\mathbf{p,q}t}\propto \delta _{\Delta \varepsilon }(\varepsilon
-\varepsilon _{ex})e^{-(ql_{ex})^{2}}w_{t}.
\end{equation}%
Here $\delta _{\Delta \varepsilon }(\varepsilon -\varepsilon _{ex})$ is the
peak energy distribution with the half-width $\Delta \varepsilon $ placed at
the energy $\varepsilon _{ex}$, $l_{ex}$ is the in-plane scale of
excitation, and $w_{t}$ is the above-introduced form-factor. After the
integration over $\mathbf{q}$-plane, the distribution $f_{\mathbf{p,x}t}$
takes the form: %A2
\begin{equation}
f_{\mathbf{p,x}t}\propto \delta _{\Delta \varepsilon }(\varepsilon
-\varepsilon _{ex})\int_{-\infty }^{t}dt^{\prime }w_{t^{\prime }}\exp \left[
-\frac{1}{2}\left( \frac{\mathbf{x}_{t-t^{\prime }}}{l_{ex}}\right) ^{2}%
\right] ,
\end{equation}%
so $f_{\mathbf{p,x}t}>0$ because of the positive functions under integral.
At $t\gg \tau _{ex}$ one obtains the distribution as a moving Gaussian peak: 
$f_{\mathbf{p,x}t}\propto \delta _{\Delta \varepsilon }(\varepsilon
-\varepsilon _{ex})\exp [-(1/2)(\mathbf{x}_{t}/l_{ex})^{2}]$. The explicit
expressions for $f_{p,x}^{\Vert ,\perp }$ introduced in analogy to Eq. (12)
can be written through the probability integrals and they have a single-peak
behavior.

Restricting ourself to a narrow energy distribution, $\Delta \varepsilon \ll
\varepsilon _{ex}$, and taking the integrals over $\mathbf{p}$-plane
according the definition (14), one obtains the concentration distribution as
follows: %A3
\begin{eqnarray}
n_{xt} \propto \exp \left[ -\frac{1}{2}\left( \frac{x}{l_{ex}}\right) ^{2}%
\right] \int_{-\infty }^{t/\tau _{ex}}d\tau e^{-2\tau ^{2}}  \nonumber \\
\times e^{-\alpha (t/\tau _{ex}-\tau )^{2}}I_{0}\left[ \alpha \frac{x}{l_{ex}%
}\left( \frac{t}{\tau _{ex}}-\tau \right) \right] ,
\end{eqnarray}%
where $I_{0}(z)$ is the zero-order Bessel function of imaginary argument.
Here we have introduced the dimensionless parameter $\alpha =v_{ex}\tau
_{ex}/l_{ex}$ with $v_{ex}=\sqrt{2\varepsilon _{ex}/m}$. Within the above
approximation, the energy distribution is given by $\mathcal{E}_{xt}\simeq
\varepsilon _{ex}n_{xt}$. Similar to Eqs. (19, 20) one obtains the flow
density $\mathbf{i}_{\mathbf{x}t}=(\mathbf{x}/|\mathbf{x}|)I_{xt}$, where
the scalar function $I_{xt}$ is written as follows: 
\begin{eqnarray}
I_{xt} \propto v_{ex}\exp \left[ -\frac{1}{2}\left( \frac{x}{l_{ex}}\right)
^{2}\right] \int_{-\infty }^{t/\tau _{ex}}d\tau e^{-2\tau ^{2}}  \nonumber \\
e^{-\alpha (t/\tau _{ex}-\tau )^{2}}I_{1}\left[ \alpha \frac{x}{l_{ex}}%
\left( \frac{t}{\tau _{ex}}-\tau \right) \right] .
\end{eqnarray}%
where $I_{1}(z)$ is the first-order Bessel function of imaginary argument.

Performing a simple numerical integration of Eqs. (A3) and (A4) one obtains
the concentration and flow distributions versus the dimensionless coordinate
and time, $x/l_{ex}$ and $t/\tau _{ex}$, as it is shown in Fig. 6 for $%
\alpha =6$. Since there are no oscillation of the classical distribution
(A2), the mean values appear to be spread monotonically.

\begin{acknowledgments}
This work has been supported in part by Ministerio de Educaci\'{o}n y
Ciencia (Spain) and FEDER under the project FIS2005-01672, and by FRSF of
Ukraine (grant No.16/2).
\end{acknowledgments}


\begin{thebibliography}{99}
\bibitem{13x} Note, that the particle conservation law, 
\[
\frac{d}{dt}\left( 2\sum_{\mathbf{p}}f_{\mathbf{p,q}=0t}+n_{t}\right) =0, 
\]%
can be verified under calculation of the time derivative of $2\sum_{\mathbf{p%
}}f_{\mathbf{p,q}=0t}$ with the use of Eqs. (8, 9).
\end{thebibliography}
\end{document}